\documentclass{jetpl}

\usepackage{xcolor}

\twocolumn

\lat

\title{Enhanced Second\d Harmonic Generation with Structured Light\\ in AlGaAs Nanoparticles Governed by Magnetic Response}

\rtitle{Enhanced Second\d Harmonic Generation with Structured Light\ldots}

\sodtitle{Enhanced second\d harmonic generation by structured light in Mie\d resonant nanoparticles}

\author{Elizaveta~V.\,Melik\D Gaykazyan$^{+*}$, Kirill~L.\,Koshelev$^{*\times}$, Jae\D Hyuck~Choi$^\circ$, Sergey~S.\,Kruk$^*$, Hong\D Gyu~Park$^\circ$, Andrey~A.\,Fedyanin$^+$, Yuri~S.\,Kivshar$^*$\/\thanks{e-mail: ysk@internode.on.net}\\~\\}

\rauthor{E.\,V.\,Melik\D Gaykazyan, K.\,L.\,Koshelev, J.\D H.\,Choi et al.}

\sodauthor{Melik\D Gaykazyan, Koshelev, Choi, Kruk, Park, Fedyanin, Kivshar}

\address{$^+$Faculty of Physics, Lomonosov Moscow State University,
119991 Moscow, Russia\\~\\
$^*$Nonlinear Physics Centre, Australian National University, ACT 2601, Canberra, Australia\\~\\
$^\times$ITMO University, 197101 Saint Petersburg, Russia\\~\\
$^\circ$Department of Physics, Korea University, 02841 Seoul, Republic of Korea\\~\\}

\dates{21 November 2018}{*}

\abstract{We employ structured light for the second-harmonic generation from subwavelength AlGaAs nanoparticles that support both electric and magnetic multipolar Mie resonances. The vectorial structure of the pump beam allows addressing selectively Mie-resonant modes and control the strength of the generated nonlinear fields.  We observe experimentally the enhancement of the second-harmonic generation for the azimuthally polarized vector beams near magnetic dipole resonance, and match our observations with the numerical decomposion of
the Mie-type multipoles for the fundamental and generated second-harmonic fields.}

\begin{document}

\maketitle

\subparagraph{1. Introduction.} The recently emerged field of dielectric Mie-resonant meta-optics is driven by the studies of dielectric nanoparticles with a high refractive index, and it shapes new active research directions of nanoscale resonant nanophotonics and dielectric nanoantennas~\cite{1Kruk2017,kivshar}. All-dielectric meta-optics employs subwavelength dielectric Mie-resonant nanoparticles as "meta-atoms"\ for creating highly efficient optical metasurfaces and metadevices~\cite{2Zheludev2012}.  The term "meta-optics"\ is usually employed to emphasize the importance of the optically-induced magnetic response of subwavelength dielectric structures which are expected to complement or even replace plasmonic components for potential applications at the nanoscale. 

High-index dielectric nanoparticles can support both electric and magnetic Mie-type resonances in the visible and near-infrared spectral ranges, and they can easily be tailored by the nanoparticle geometry. The Mie-resonant silicon nanoparticles have recently received considerable attention for applications in nanophotonics and metamaterials \cite{1Kruk2017} including optical nanoantennas, wavefront-shaping metasurfaces, and nonlinear frequency generation. Importantly, the resonant excitation of strong multipolar Mie-type localized modes may also lead to the enhancement of electric and magnetic fields in such structures and drive {\em novel nonlinear effects}.

Second-harmonic (SH) generation is one of the most important nonlinear processes in optics. For this parametric process, the frequency of an incident light beam is doubled inside of a nonlinear crystal. SH generation is employed for many applications, including laser sources and nonlinear spectroscopy. Remarkably, harmonic generation processes at the nanoscale are governed by the resonant properties of nanoscale structures and the mode volume being independent on the phase-matching principle.

In the last few years, the role of nonlinear optically-induced magnetic response was intensively addressed in nanophotonics. In particular, Shcherbakov {\em et al.}~\cite{3Shcherbakov2014} and Melik-Gaykazyan {\em et al.}~\cite{melik} employed silicon nanoantennas and demonstrated that the nonlinear response from silicon nanodisks prevails over the harmonics generated from the bulk silicon, and it becomes possible to reach conversion efficiency high enough for the generated visible light to be observed by naked eye.  Generation of different localized Mie-resonant modes can reshape completely the physics of nonlinear effects at the nanoscale \cite{4Smirnova2016}. Utilizing the Mie resonances in dielectric nanoparticles has recently been recognized as a promising strategy to gain higher efficiency of nonlinear parametric processes at low modal volumes, and achieve novel functionalities originating from optically-induced nonlinear magnetism \cite{5Kruk2017}.  Generated multipoles are closely related to the vectorial nature of the pump beam, and they can be distinguished in the far-field region by their polarization and modal decomposition~\cite{5Kruk2017}.

Focused and structured vector beams have already been employed for the harmonic generation, and they have been used in nanophotonics as a versatile tool, in comparison with the linearly polarized light, for all-optical characterization. A comprehensive review of microscopic techniques based on nonlinear optical processes in nanoscale structures and capabilities of nonlinear microscopy has been discussed recently by Bautista and Kauranen~\cite{6Bautista2016} who demonstrated that the vectorial focusing of linear and circular polarized (conventional) and radial and azimuthal polarized (unconventional) light can spread the range of capabilities for nonlinear microscopy techniques.

Recently, we have studied the third-harmonic generation by structured (radially and azimuthally polarized) light \cite{7Melik2017} and, by tailoring the vectorial structure of the pumping light, we have demonstrated a control of both strength and polarization of the excited harmonic fields, also addressing selectively different types of multipolar Mie resonances. 

In this Letter, we employ our earlier approach for third-harmonic-generation experiments and study, for the first time to our knowledge, nonlinear second-harmonic spectroscopy of individual Mie-resonant AlGaAs nanoparticles excited with structured light. Earlier, such problems have been considered only for plasmonic nanoparticles and nanoparticle oligomers~\cite{6Bautista2016,OLT2017,bautista2018}. We observe the substantial enhancement of the second-harmonic nonlinear signal generated by an azimuthally polarized pump beam. We supplement our observations by numerical mode-decomposion analysis and reveal the contribution of the modes at the double frequencies, as well as confirm that structured light can be employed as an efficient tool for the enhanced harmonic generation and subwavelength mode control at the nanoscale. 

\subparagraph{2. Numerical approach.} We simulate numerically both linear and nonlinear optical response, and eigenmode spectra of a single Al$_{0.2}$Ga$_{0.8}$As nanodisk using three-dimensional electromagnetic simulations implemented with the finite-element method in COMSOL Multiphysics. The nanodisk is placed on a slide glass substrate, all material properties including real losses are imported from the tabulated data \cite{8refractiveindex}. The vector structure of a pump beam is simulated using the nonparaxial representation \cite{11Chaumet2006}. We modify the standard formulas to include rigorously the effect of a substrate by adding reflected and transmitted plane waves into the nonparaxial expansion. For calculations, we take the beam waist radius equal to $1.5$ um. 

\begin{figure}[ht]
\centering
\includegraphics[width=1\columnwidth]{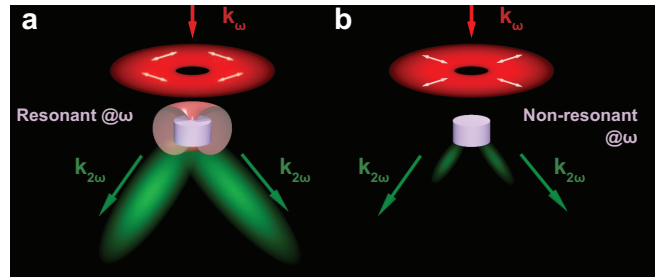}
\caption{Fig.1. (Color online) Concept image of the second-harmonic (SH) generation in subwavelength nanoparticles with structured light. (a) Enhanced SH signal from a particle resonantly excited at the fundamental wavelength by an azimuthally polarized pump beam. (b) SH signal from a non-resonant at the fundamental frequency disk excited by a radially polarized vector beam. White arrows depict the polarization distribution in the cross-section of a cylindrical vector pump beam.}
\label{F1}
\end{figure}

\begin{figure*}[ht]
\centering
\includegraphics[width=1.7\columnwidth]{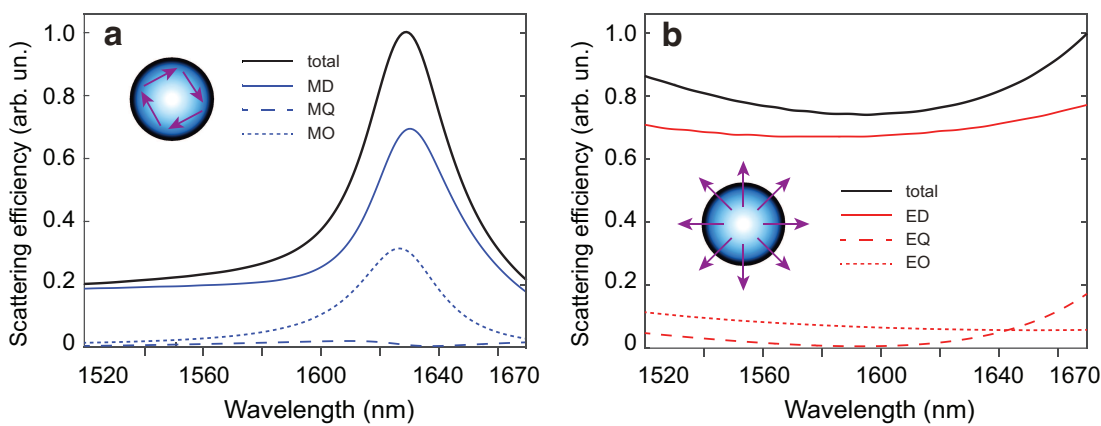}
\caption{Fig.2. (Color online) Numerical results. Multipolar decomposition of the scattering efficiency spectrum of a single Al$_{0.2}$Ga$_{0.8}$ disk (height is $650$~nm, diameter is $935$~nm) excited by a vector beam of (a)~azimuthal and (b)~radial polarization. Solid black line represents the total sum of multipolar contributions. Insets show the amplitude and polarization of electric field of the incident beam in the central cross section of the disk. The total scattering efficiency is independently normalized on the maximal value for each case. ED and MD, electric and magnetic dipoles; EQ and MQ, electric and magnetic quadrupoles; EO and MO, electric and magnetic octupoles.}
\label{F2}
\end{figure*}

We choose geometrical parameters of an AlGaAs resonator to excite a magnetic multipolar resonance at the fundamental wavelength by an azimuthally polarized pump beam in the considered spectral range of near IR (see Fig.~\ref{F1}a). The corresponding geometry of excitation by a radially polarized pump beam is shown in Fig.~\ref{F1}b. The height of a disk is equal to 650~nm, its diameter is 935~nm. By decreasing the disk radius one can shift this resonance spectral position to the blue range of the spectrum. Figure~\ref{F2} shows scattering efficiency spectra of an AlGaAs nanodisk of the specified parameters in the linear regime. One can see that indeed the azimuthal polarization excites magnetic Mie-type multipolar modes, whilst the contribution of electric multipoles only occurs for the radial excitation case, which is a typical tendency for subwavelength high-index particles \cite{7Melik2017,14Das2015,das2017}. Contrary to the azimuthally polarized excitation, there are not spectral features for the disk pumped by a radially polarized cylindrical vector beam.

The nonlinear optical response is estimated within the undepleted pump approximation \cite{9Boyd}. The induced second-order nonlinear susceptibility tensor $\chi^{(2)}$ corresponds to zincblende crystalline structure of AlGaAs, determining the value of induced polarization $\mathbf{P}^{(2\omega)}$ at the SH frequency

\begin{equation}
P^{(2\omega)}_i=\varepsilon_0\chi^{(2)}_{ijk}E^{(\omega)}_j E^{(\omega)}_k,    
\end{equation}
where $\mathbf{E}^{(\omega)}$ is the full electric field inside the resonator at the fundamental frequency. We assume the value of non-zero $\chi^{(2)}$ components equal to 290~pm/V \cite{10Shoji1997}.

\begin{figure*}
\centering
\includegraphics[width=1.7\columnwidth]{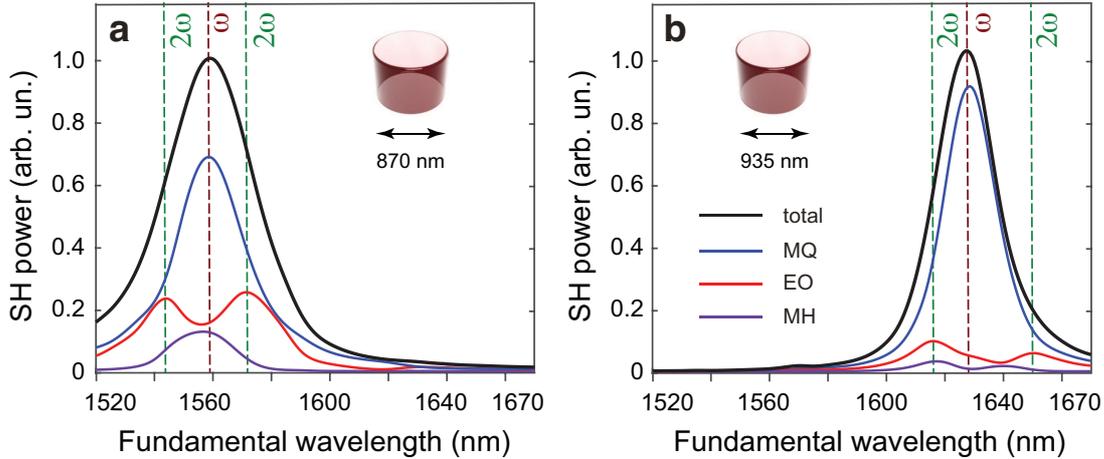}
\caption{Fig.3. (Color online) {\bf Numerical results.} Multipolar decomposition of the second-harmonic (SH) field generated by an azimuthal incident beam in an Al$_{0.2}$Ga$_{0.8}$As nanodisk with height 650~nm and diameter (a)~870ñm and (b)~935~nm, respectively. Notations are the following: MQ, a magnetic quadrupole; EO, an electric octupole; MH, a magnetic hexadecapole. Red and green dashed vertical lines mark the positions of nanodisk eigenmodes at the fundamental and SH frequencies, respectively. The total SH power is independently normalized on the maximal value for each case.}
\label{F3}
\end{figure*}

We perform the nonlinear numerical simulations of the generated SH electromagnetic fields, their multipolar decomposition and disk eigenmode analysis (Fig.~\ref{F3}). It should be pointed out that the spectral position of eigenmodes does not depend on the type of the external excitation and is defined by the parameters of the sample only. The dominant contribution to the SH enhancement of the particle pumped by azimuthally polarized light is made by the mode at the fundamental frequency. Moreover, the spectral features of the electric octupolar (EO) contribution can be interpreted by disk eigenmode analysis at the double pump frequency. We present the nonlinear calculation results for two different values of the radius to confirm the spectral shifts of optical modes and their dependence on resonator sizes.

Both scattering spectra and SH power spectra are associated with the electric and magnetic multipolar contributions calculated in the spherical representation \cite{12Grahn2012}. The conventional multipolar analysis does not take the effect of the substrate reflection into account, which provides an error in absolute values of the total contribution of multipoles of order of several percents. In both linear and nonlinear multipolar calculations, the total power is calculated over the full solid angle.

\subparagraph{3. Experimental results and discussion.} Nanoparticles are made of a custom-designed wafer consisted of a GaAs substrate, a 75-nm sacrificial layer of Al$_{0.51}$In$_{0.49}$P and a 650-nm layer of Al$_{0.2}$Ga$_{0.8}$As with a crystalline axis [100]. The etch mask is created by the method of electron beam lithography; then, sequential processes of dry etching to form pillars and selective wet-etching of the sacrificial layer in solution of HCl are applied. AlGaAs disks are picked up from the GaAs substrate and dropped down to a slide glass substrate. The final dimensions of a nanopillar are checked by scanning electron microscopy. The nanodisk diameter is found to be 935~nm,  in correspondence with the value we used for numerical calculations.

In order to conduct nonlinear spectroscopy measurements, we construct an experimental setup based on an optical parametric amplifier (Hotlight Systems, MIROPA-fs-M, pulse duration of about 300~fs) pumped by a pulsed Yb laser by High Q Laser GmbH with a repetition rate of 21~MHz and operating wavelength of 1043~nm. The considered spectral range of the pump is 1520-1670~nm. We implement a silicon-based q-plate metasurface~\cite{13Kruk2016} to create the vectorial structure of a pump beam from a linearly polarized Gaussian laser beam. The $q$-plate is placed in a telescopic system of two lenses (Thorlabs AC254-200-C-ML and AC254-050-C-ML with focal distances of 20 and 5~cm, respectively). By flipping linear polarization from horizontal to vertical using an achromatic half-wave plate (Thorlabs AHWP05M-1600) we switch the pump beam type from azimuthal to radial. There is a longpass infrared filter (Thorlabs FELH1300) after the $q$-plate and before the sample to clear up the spectrum of a pump beam. The polarization type of a pump beam is checked by an analyzer which is a combination of an achromatic quarter wave-plate and broadband wire-grid polarizer -- Thorlabs AQWP05M-1600 and WP25M-UB, respectively. The pump beam is observed by a near-infrared InGaAs camera Xenics Bobcat-320 with a 150-mm focal distance achromatic doublet (AC508-150-C-ML). A train of wavelength tunable femtosecond laser pulses is focused from the front side of the sample mounted on a three-dimensional stage to a beam waist size, which is close to a diffraction limit, by a Mitutoyo MPlanApo~NIR objective lens, 100x infrared, 0.70~numerical aperture (NA). Therefore the fabricated nanoparticles are considered as isolated due to the spacing value of 10~$\mu$m. The SH signal is collected by an Olympus objective lens MPlanFL~N (100x visible, 0.90~NA) and detected by a visible cooled CCD camera (Starlight Xpress Ltd, Trius-SX694) with a 150-mm focal distance achromatic doublet (AC508-150-A-ML). The detected signal is filtered out by a set of filters (a~colored glass bandpass filter Thorlabs FGS900 and UV fused silica filter with dielectric coating FELH0650). The SH signal is normalized over a spectral function of the setup which includes filter transmittance, laser power, and detector sensitivity spectra. The origin of the SH signal is verified by the direct measurement of its spectrum (by a visible spectrometer Ocean Optics QE~Pro) and its power dependence which is in a quadratic manner with a saturation average power level of about 250~$\mu$W.

\begin{figure}[ht]
\centering
\includegraphics[width=1.0\columnwidth]{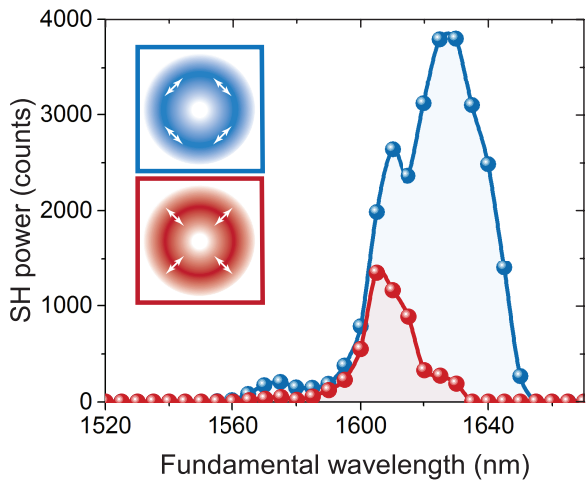}
\caption{Fig.4. (Color online) Experimental results. Second-harmonic generation spectroscopy of a single Al$_{0.2}$Ga$_{0.8}$As resonator with a diameter of 935~nm pumped by structured light. Blue dotes correspond to the azimuthally polarized pump beam, red dots -- to the radial excitation, as shown in the insets. White arrows illustrate the polarization distributions in the central cross-section of cylindrical vector pump beams.}
\label{F4}
\end{figure}

Our experimental results on the SH generation spectroscopy of an individual AlGaAs resonator of the specific geometrical parameters with cylindrical vector beams are shown in Fig.~\ref{F4}. As is expected, we observe the enhancement of the SH signal collected by an objective lens (0.9~NA) for the case of a particle which is resonant at the fundamental wavelength and pumped by an azimuthally polarized laser light. The additional local peaks of the experimental SH spectrum are associated with the excitation of disk eigenmodes at the SH wavelengths, particularly, the increased contribution of an EO mode which could be understood by taking into account the imperfections of fabricated resonators as well as changes of a focused pump beam waist diameter by tuning the laser wavelength (while it is considered as a constant in our simulations). We observe the four-time less enhancement of the nonlinear response provided by a non-resonant in the linear regime particle excited by a radially polarized laser beam; the spectral position of this enhancement corresponds to the calculated eigenmode of the disk at the double fundamental frequency. This result provides a potential tool of the use of radially polarized light to probe the resonances at doubled frequencies for such type of nanoresonators.

\subparagraph{6. Conclusions.} We have employed cylindrical vector beams to enhance the second-harmonic generation response of a single AlGaAs nanoresonator and have  conducted spectroscopy measurements. We have proposed the theoretical interpretation of the observed spectral features of the nonlinear signal generation based on the calculations of eigenmodes of the nanoparticle excited at the fundamental and double frequencies. Our findings suggest the potential of subwavelength semiconductor nanoresonators to become the basis of advanced nanophotonic devices.  

The authors thank Prof. Barry Luther\d Davies for support. E.V.M. acknowledges the funding from the Student Mobility Scholarship of the President of the Russian Federation, the Russian Ministry of Education and Science (\#14.W03.31.0008), and the Russian Foundation for Basic Research (the~project No.~18-32-01038). A.A.F. acknowledges the support of the Russian Foundation of Basic Research (project \#18-29-20097). Nonlinear simulations were supported by the Russian Science Foundation (grant 18-72-10140). K.L.K. acknowledges a support from the Foundation for the Advancement of Theoretical Physics and Mathematics ”BASIS” (Russia).


\begin{thebibliography}{99}

\bibitem{1Kruk2017}
S.~Kruk and Yu.\,S.~Kivshar, Functional meta-optics and nanophotonics governed by Mie resonances, ACS Photonics {\bf 4}, 2638 (2017).

\bibitem{kivshar}
Y.\,S.~Kivshar, All-dielectric meta-optics and non-linear nanophotonics, National Science Review {\bf 5}, 144-158 (2018). 

\bibitem{2Zheludev2012}
N.~Zheludev and Yu.\,S.~Kivshar, From metamaterials to metadevices, Nature Materials {\bf 11}, 917924 (2012).

\bibitem{3Shcherbakov2014}
M.\,R.~Shcherbakov, D.\,N.~Neshev, B.~Hopkins, A.\,S.~Shorokhov, I.~Staude, E.\,V.~Melik-Gaykazyan, M.~Decker, A.\,A.~Ezhov, A.\,E.~Miroshnichenko, I.~Brener, A.\,A.~Fedyanin, and Y.\,S.~Kivshar, Enhanced third-harmonic generation in silicon nanoparticles driven by magnetic response, Nano Lett. {\bf 14}, 6488 (2014).

\bibitem{melik}
E.\,V.~Melik-Gaykazyan, M.\,R.~Shcherbakov, A.\,S.~Shorokhov, I.~Staude, I.~Brener, D.\,N.~Neshev. Yu.\,S.~Kivshar, and A.\,A.~Fedyanin, Third-harmonic generation from Mie-type resonances of isolated all-dielectric nanoparticles, Phil. Trans. R. Soc. A {\bf 375}, 20160281 (2017). 

\bibitem{4Smirnova2016}
D.~Smirnova and Y.\,S.~Kivshar, Multipolar nonlinear nanophotonics, Optica {\bf 3}, 1241 (2016). 

\bibitem{5Kruk2017}
S.~Kruk, R.~Camacho-Morales, L.~Xu, M.~Rahmani, D.\,A.~Smirnova, L.~Wang, H.\,H.~Tan, C.~Jagadish, D.\,N.~Neshev, and Y.\,S.~Kivshar, Nonlinear optical magnetism revealed by second-harmonic generation in nanoantennas, Nano Lett. {\bf 17}, 3914 (2017). 

\bibitem{6Bautista2016}
G.~Bautista and M.~Kauranen, Vector-field nonlinear microscopy of nanostructures, ACS Photonics {\bf 3}, 1351 (2016). 

\bibitem{7Melik2017}
E.\,V.~Melik-Gaykazyan, S.\,S.~Kruk, R.~Camacho-Morales, L.~Xu, M.~Rahmani, K.~Zangeneh Kamali, A.~Lamprianidis, A.\,E.~Miroshnichenko, A.\,A.~Fedyanin, D.\,N.~Neshev, and Y.\,S.~Kivshar, Selective third-harmonic generation by structured light in Mie-resonant nanoparticles, ACS Photonics {\bf 5}, 728 (2018).

\bibitem{OLT2017}
A.~Ahmadivand, R.~Sinha, and N.~Pala, Magnetic Fano resonances in all-dielectric nanocomplexes under cylindrical vector beams excitation, Optics \& Laser Technology {\bf 90}, 65-70 (2017). 

\bibitem{bautista2018}
G.~Bautista, C.~Dreser, X.~Zhang, D.\,P.~Kern, M.~Kauranen, and M.~Fleischer, Collective effects in second-harmonic generation from plasmonic oligomers, Nano Lett. {\bf 18}, 2571 (2018). 


\bibitem{8refractiveindex}
http://refractiveindex.info/.

\bibitem{11Chaumet2006}
P.\,C.~Chaumet, Fully vectorial highly nonparaxial beam close to the waist, J. Opt. Soc. Am. A {\bf 23}, 3197 (2006).

\bibitem{14Das2015}
T.~Das, P.\,P.~Iyer, R.\,A.~DeCrescent, J.\,A.~Schuller, Beam engineering for selective and enhanced coupling to multipolar resonances, Phys. Rev. B {\bf 92}, 241110 (2015).

\bibitem{das2017}
T.~Das and J.\,A.~Schuller, Dark modes and field enhancement in dielectric dimers illuminated by cylindrical vector beams, Phys. Rev. B {\bf 95}, 201111 (2017). 

\bibitem{9Boyd}
R.\,W.~Boyd, {\it Nonlinear Optics}, Elsevier (2003).

\bibitem{10Shoji1997}
I.~Shoji, T.~Kondo, A.~Kitamoto, M.~Shirane, and R.~Ito, Absolute scale of second-order nonlinear-optical coefficients, J. Opt. Soc. Amer. B {\bf 14}, 2268 (1997).

\bibitem{12Grahn2012}
P.~Grahn, A.~Shevchenko, and M.~Kaivola, Electromagnetic multipole theory for optical nanomaterials, New J. Phys. {\bf 14}, 093033 (2012).

\bibitem{13Kruk2016}
S.~Kruk, B.~Hopkins, I.\,I.~Kravchenko, A.~Miroshnichenko, D.\,N.~Neshev, Y.\,S.~Kivshar, Broadband highly efficient dielectric metadevices for polarization control, APL Photonics {\bf 1}, 030801 (2016).


\end{thebibliography}
\end{document}